\documentclass[conference]{IEEEtran}
\IEEEoverridecommandlockouts
\usepackage{cite}
\usepackage{amsmath,amssymb,amsfonts}
\usepackage{algorithmic}
\usepackage{graphicx}
\usepackage{textcomp}
\usepackage{xcolor}
\usepackage{float}
\restylefloat{table}
\usepackage{hyperref}
\usepackage{graphbox} 

\usepackage{subcaption}
\def\BibTeX{{\rm B\kern-.05em{\sc i\kern-.025em b}\kern-.08em
    T\kern-.1667em\lower.7ex\hbox{E}\kern-.125emX}}
\begin{document}

\title{Learning Self-Supervised Representations for Label Efficient Cross-Domain Knowledge Transfer on Diabetic Retinopathy Fundus Images
}

\author{
    \IEEEauthorblockN{
        Ekta Gupta\textsuperscript{$\gamma$},
        Varun Gupta\textsuperscript{$\gamma$},
        Muskaan Chopra\textsuperscript{$\gamma$},
        Prakash Chandra Chhipa\textsuperscript{$\delta$} and
        Marcus Liwicki\textsuperscript{$\delta$}\\
        \textit{\textsuperscript{$\delta$} Machine Learning Group, EISLAB, Lule\aa~Tekniska Universitet, Lule\r{a}, Sweden}\\
        \textit{\{prakash.chandra.chhipa, marcus.liwicki\}@ltu.se}\\
        \textit{\textsuperscript{$\gamma$} Chandigarh College of Engineering and Technology, Punjab University, Chandigarh, India}\\
        \textit{ekta\_cse@ccet.ac.in, varungupta@ccet.ac.in, chopramuskaan47@gmail.com}\\
     }
    }


\maketitle
\begin{abstract}
This work presents a novel label-efficient self-supervised representation learning-based approach for classifying diabetic retinopathy (DR) images in cross-domain settings. Most of the existing DR image classification methods are based on supervised learning which requires a lot of time-consuming and expensive medical domain experts-annotated data for training. The proposed approach uses the prior learning from the source DR image dataset to classify images drawn from the target datasets. The image representations learned from the unlabeled source domain dataset through contrastive learning are used to classify DR images from the target domain dataset. Moreover, the proposed approach requires a few labeled images to perform successfully on DR image classification tasks in cross-domain settings. The proposed work experiments with four publicly available datasets: EyePACS, APTOS 2019, MESSIDOR-I, and Fundus Images for self-supervised representation learning-based DR image classification in cross-domain settings. The proposed method achieves state-of-the-art results on binary and multi-classification of DR images, even in cross-domain settings. The proposed method outperforms the existing DR image binary and multi-class classification methods proposed in the literature. The proposed method is also validated qualitatively using class activation maps, revealing that the method can learn explainable image representations. The source code and trained models are published on GitHub\footnote{\href{https://github.com/prakashchhipa/Learning-Self-Supervised-Representations-for-Label-Efficient-Cross-Domain-Knowledge-Transfer-on-DRF} {https://github.com/prakashchhipa/Learning-Self-Supervised-Representations-for-Label-Efficient-Cross-Domain-Knowledge-Transfer-on-DRF}}.

\end{abstract}


\section{Introduction}
In the medical imaging area, artificial intelligence (AI), a topic characterized broadly by the building of computerized systems capable of doing tasks \cite{b1} \& \cite{b2} that ordinarily require human intelligence, has significant potential. Automated radiology workflows have significantly benefited from machine learning and deep learning techniques. Although AI models have the potential to revolutionize clinical practice, they have been hampered by significant implementation and regulatory obstacles \cite{b3}. Almost all constraints may be traced back to a major issue: a dearth of medical image data to train and test AI algorithms \cite{b4}. The generalizability and accuracy of developed solutions are hampered because most research institutions and enterprises only have limited access to annotated medical images. Large datasets, including high-quality images and annotations, are still necessary to train, validate, and test the AI systems \cite{b5}. 
In the absence of data that has been appropriately labeled, this procedure becomes prohibitively expensive, time-demanding, and inherently unstable.
\begin{figure}[htbp]
\centerline{
\includegraphics[width=9cm, height=4cm]{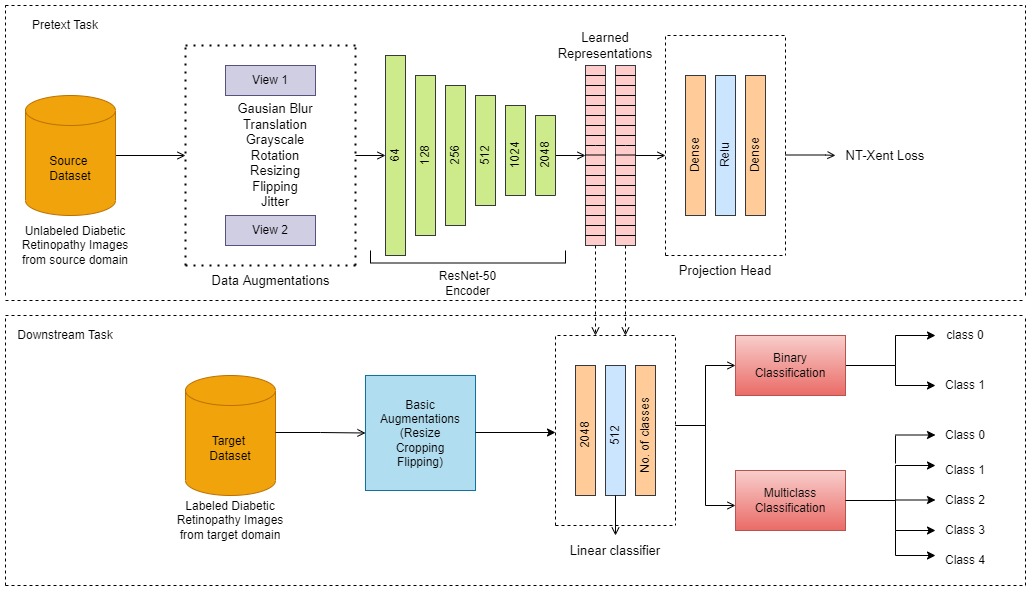} 
}
\caption{Schematic presents the contrastive learning-based self-supervised cross-domain knowledge transfer. Pretraining is performed on the source dataset (EyePACS), and downstream tasks are performed on cross-domain targets (APTOS 2019, MESSIDOR, and Fundus Images).}
\label{fig1}
\end{figure}
Labeled biomedical images are incredibly scarce, and multiple experts are required to annotate each image \cite{b6} manually. Massive amounts of health data are being generated and collected. These data range from in-hospital monitoring to wearable. Coding and annotating this data is impractical \cite{b6}\cite{b7}.  In addition, the pretrained models obtained from natural images do not apply directly to medical images since their intensity distribution is different.  Besides that, annotating natural images is simple; all that is required is basic human knowledge \cite{b8}.
Nevertheless, in-depth knowledge is necessary for the annotation of medical images. The average medical image has over a billion pixels, significantly larger than others. The annotation process is highly error-prone and expensive, and experts cannot always identify a particular feature. A potential solution is to train models on unlabeled images using self-supervised learning \cite{b9}\cite{b10}.

Most supervised learning methods require labeled data to train a machine. Unfortunately, obtaining good-quality labeled data can be cost-effective and time-consuming. Additionally, the data preparation lifecycle can be extremely long and complicated, including cleaning, filtering, annotating, reviewing, and restructuring according to a training framework \cite{b11}.
Another approach has been used to deal with scarce biological data: domain adaptation-based self-supervised learning. Self-supervised learning (SSL), an alternative to supervised learning and transfer learning, has emerged as a viable possibility \cite{b12}. While self-supervised learning is distinct from transfer learning, both rely on acquiring representations from a secondary pretext activity and the subsequent transfer of those representations to the main focus task \cite{b13}\cite{b14}. The data utilized for the pretraining phase and the downstream task might be taken from one or more separate data sources in domain adaptation self-supervised learning, unlike transfer learning \cite{b15}.

This work aims to prove that self-supervised learning can be used as a preliminary step in medical image classification. Contributions to this work are:
\begin{itemize}
\item This work proposes a domain adaptation-based self-supervised learning approach to learn image representations from diabetic retinopathy fundus images.
\item Self-supervised learning of diabetic retinopathy image representations from unlabelled datasets has been validated in cross-domain settings, as shown in Figure 1.
\item Results indicate that the proposed work outperforms the existing methods of DR classification.
\end{itemize}
\section{Related Work}
In recent years, unsupervised learning has significantly progressed (SSL).  Since it is useful for learning feature representations from image datasets without image labels, it has become a primary focus for academic investigation. Medical image classification tasks like detecting diabetic retinopathy, classifying brain age~\cite{b35}, recognizing cancer in histopathology~\cite{b34}, identifying pneumonia in X-ray~\cite{b36}, many others have shown progress using self-supervised learning methods, demonstrating state-of-the-art performance. This article focuses on self-supervised learning techniques pertaining to images of diabetic retinopathy.\\
In an approach, Truong et al.  proposed the fusion of embeddings from multiple SSL models. Then the fused embeddings are combined with self-attention for feature learning. However, it did not use domain adaptation, as the datasets used in the pretext and downstream tasks are the same.\\
Taleb et al. \cite{b20} developed a series of five proxy tasks: 3D contrastive predicting coding, 3D rotation prediction, 3D jigsaw puzzles, 3D patch location, and 3D exemplar networks to learn the feature representations in the pretext task. Lin et al. \cite{b21} proposed a multilabel classification method using rotation SSL with graph CNN to learn fundus images' representations. Another work by Srinivasan et al. \cite{b22} trained a ResNet50 model using the MoCo-V2 approach in the pretext task to classify diabetic retinopathy images in the downstream task. The authors used a similar dataset for both the pretext and downstream tasks.  
Yap and Ng \cite{b23} proposed a contrastive learning framework to create a patch-level dataset for pretext tasks by extracting the class activation maps from the labeled and unlabelled datasets.

\subsection{SSL methods for DR segmentation}
Segmentation of diabetic retinopathy using self-supervised learning has been explored partially. Tian et al. \cite{b18} proposed a multi-class Strong Augmentation via Contrastive Learning (MSACL) approach for detecting unsupervised anomalies. The author also proposes a contrastive loss that combines contrastive learning with multi-centered loss to cluster the samples of the same class. These unsupervised models need to be well-trained. Otherwise, they can learn ineffective image representations. Another author, Kukacka et al. \cite{b19}, also proposed an approach for lesion segmentation by pretraining a U-net encoder in the pretext task.

\subsection{Reconstruction-based SSL methods for DR classification}
Many efforts have been made for the diabetic retinopathy classification task using reconstruction-based self-supervised learning methods. Holmberg et al. \cite{b16} proposed a cross-domain U-net-based system to generate the retinal thickness used for the classification during the downstream task. Other authors, Ngyyen et al., learned the features of the target dataset by using a self-supervised contrastive learning method on reconstructed retinal images. This work reconstructs images and features learning on the target dataset. In the proposed work, representations learning is performed on the source dataset of diabetic retinopathy and applied to those learned representations to a different dataset of diabetic retinopathy. In addition, a few authors also proposed a multi-modal-based reconstruction of images. Hervella et al. \cite{b3} performed multimodal reconstruction using U-Net for the segmentation task of the optic disk and cup in retinography and angiography images, and Li et al. \cite{b17} trained a CycleGAN model on the source dataset to learn the mapping function between the images and also learned both the modality-invariant and patient-similarity features in the pretext task. One more work by Cai et al. proposed a transformer-based framework in combination with a multitask decoder to learn the representations of the reconstructed images. 
Most works discussed above used adversarial learning methods to reconstruct the images. However, these methods provide inferior performance or have unstable training. Representation learning is pixel-based learning in the existing reconstruction-based methods, but the work focuses on learning representations at the visual concept level.

As was seen in the preceding review of the relevant literature, most existing SSL approaches employ the same dataset for both the pretext task in the source domain and the downstream task in the target domain. Progress has been seen in the knowledge transfer field, but no one has extensively explored the domain adaptation. The proposed work concretely focuses on cross-domain contrastive learning. To identify DR images from a different domain, this study presents an SSL strategy to reuse the representations learnt on one unlabeled dataset from the source domain during the pretext job.

\section{Diabetic Retinopathy Dataset Description}
Diabetic retinopathy is a major cause of blindness among people of working age in developed countries. It is a prevalent eye disease that affects more than 93 million people globally. Diabetic retinopathy detection is currently a time-consuming and laborious technique that requires a skilled person to analyze and interpret digital color fundus images of the retina. The public datasets for diabetic retinopathy are:

\subsection{Subset of EyePACS}
Eye disease Diabetic Retinopathy (DR)\footnote{\href{https://www.kaggle.com/c/diabetic-retinopathy-detection/data}{https://www.kaggle.com/c/diabetic-retinopathy-detection/data}} is linked to long-term diabetes. If DR is caught early enough, visual loss can be halted. A comprehensive collection of high-resolution retina images captured using various imaging settings are accessible. Every subject has both a left and right field available to them. Images are identified not just with a subject id but also as being on the left or the right. A medical professional has determined diabetic retinopathy on a scale of 0 to 4.
\subsection{APTOS 2019}
Numerous people are affected by diabetic retinopathy, the most common reason for vision loss among adults in their 40s and 50s. Aravind Eye Hospital can help people in rural areas without easy access to medical screening in India's efforts to find and prevent this condition there. The answers will be available to other Ophthalmologists through the 4th Asia Pacific Tele-Ophthalmology Society (APTOS) Symposium\footnote{\href{https://www.kaggle.com/competitions/aptos2019-blindness-detection/data}{https://www.kaggle.com/competitions/aptos2019-blindness-detection/data}}. A vast collection of retina images was collected using fundus photography in various situations made available. A clinical expert has determined that each image has been graded for its severity on a scale of 0 to 4.

\subsection{MESSIDOR-I}
Diabetic retinopathy detection is now a labor-intensive and time-consuming method that requires a qualified doctor to use digital color fundus images of the retina. It is known as MESSIDOR (Methods to Evaluate Segmentation and Indexing Techniques in the Field of Retinal Ophthalmology in French)\footnote{\href{https://www.adcis.net/en/third-party/messidor}{https://www.adcis.net/en/third-party/messidor}} \cite{b24}. The retinopathy grades are determined on a scale of 0 to 3.

\subsection{Fundus Images}
The Department of Ophthalmology provided the 757 color fundus images~\cite{b36} included in this collection from the Hospital de Clnicas, Facultad de Ciencias Médicas, Universidad Nacional de Asunción, Paraguay. The Zeiss brand's Visucam 500 camera was utilized for the process of acquiring the retinographies. Fundus images have been classified into 7 distinct groups on a scale of 1 to 7.
\begin{table}[htbp]
\caption{Diabetic retinopathy dataset description}
\begin{center}
\begin{tabular}{|c|c|c|}
\hline
\textbf{\textit{Dataset}}& \textbf{\textit{Total Images}}& \textbf{\textit{No. of Classes}} \\
\hline
\textit{Subset of EyePACS}& \textit{31615}& \textit{5} \\
\hline
\textit{APTOS 2019}& \textit{3660}& \textit{5} \\
\hline
\textit{MESSIDOR-I}& \textit{1200}& \textit{4} \\
\hline
\textit{Fundus Images}& \textit{747}& \textit{7} \\
\hline
\end{tabular}
\label{tab1}
\end{center}
\end{table}
\\ Table \ref{tab1} shows the dataset description of diabetic retinopathy.
\begin{figure}[htbp]
\centerline{\includegraphics[width=9cm, height=6cm]{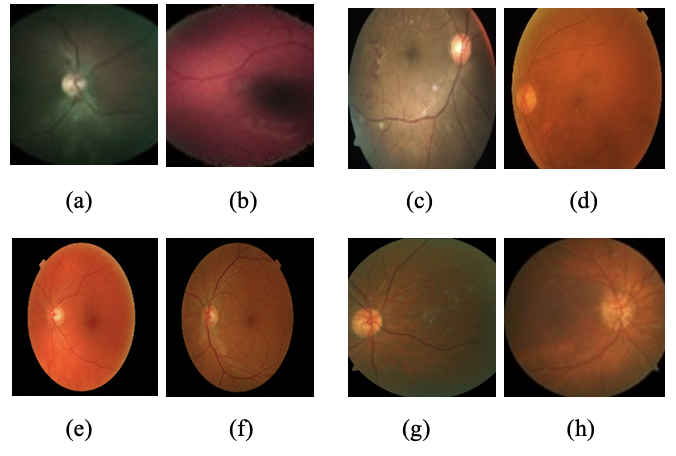} }
\caption{Sample images- (a)(b) subset of EyePACS Dataset, (c)(d) Messidor-I dataset,
(e)(f) APTOS 2019 dataset, (g)(h) Fundus images.}
\label{fig}
\end{figure}
\section{Self-supervised Cross Domain Knowledge Transfer Framework}
 The proposed framework consists of two main tasks: (i) pretext task, i.e., representation learning of images from source domain DR dataset (a subset of EyePACS) (ii) downstream task, i.e., classification of DR(Diabetic Retinopathy) images from the target domain datasets ( APTOS 2019, Messidor-I and Fundus Images). In the pretext task, the proposed approach applies various augmentations like flipping, affine transformations, jitter, grayscale, etc., to create different views from the images. The different views created from the same image act as positive pairs, and views from different images act as negative pairs. Then, image representations are learned through contrastive learning from positive and negative pairs of images. These learned representations of images act as input to the downstream task. This task does not require labeled images for representation learning as shown in Figure \ref{fig1}.\\
The downstream task involves binary as well as multi-class classification of DR images. The model pretrained for learning image representation during the pretext task act as initialization for performing the downstream task, i.e., classification of DR images. Now the downstream task requires fewer labeled images for performing DR classification.
Figure \ref{fig1} provides a detailed architecture of the proposed approach. The objective of the proposed approach is to obtain representations that are robust to domain shift and generalizable to the downstream task. The proposed approach uses an unlabeled source dataset to learn the representations and a labeled target dataset to solve the classification task by reusing these features learned from the source dataset. The representations have been learned using the SimCLR (simple framework for contrastive learning) method \cite{b25}. As discussed, positive and negative pairs of DR images are created from unlabelled DR images using different augmentations like a Gaussian blur, flipping, translation, rotation, jitter, etc. These positive and negative pairs of images are fed to the encoder network. The encoder network consists of a ResNet-50 backbone and a projection head containing two fully connected layers of 2048 and 1024 neurons, respectively. This network is trained on positive and negative views of images using Normalized Temperature-scaled Cross-Entropy (NT-Xent) as the loss function, which tries to pull positive pairs close and push away the negative pairs. This loss function is defined as: 
$$
\ell\left(\mathrm{z}_{\mathrm{i}}, \mathrm{z}_{\mathrm{j}}\right)=-\log \frac{\exp \left(\operatorname{sim}\left(Z_i Z_j\right) / T\right)}{\sum_{k=1}^{2 n} 1_{k \neq i} \exp \left(Z_i Z_k / T\right)}
\hspace{1cm}...(1) $$
Where 
$z_{i}$ and $z_{j}$ are representations of positive pairs, T is the temperature parameter, n is the number of images, and sim() represents the similarity function.
This loss function is the negative log-likelihood of similarity between positive pairs to the ratio of similarities between all possible positive and negative pairs. This loss function is a softmax function normalized using a temperature parameter.
In the downstream task, the proposed approach performs binary and multi-classification of DR images separately from the target domain datasets (Messidor-I and APTOS 2019). During this phase, primary augmentations such as resizing, flipping, and cropping are applied to the target dataset of DR images to create different views. The encoder backbone ResNet-50 weights trained for the pretext task are used as the initialization for the network being trained for the downstream task. The projection head of the network used in the pretext task is replaced with two fully connected layers, i.e. (2048, 512) and (512, no. of classes). The proposed approach performs binary and multi-classification on APTOS 2019, Messidor-I, and Fundus Images datasets during this phase.

\section{Experiments and Results}
To investigate the proposed domain adaptation framework, three datasets are explored - APTOS 2019, Messidor-I, and Fundus Image dataset. The investigation is performed in a manner that self-supervised pretrained the model on one source dataset, a Subset of EyePACS, and finetuned on the above-mentioned three target datasets. The classification task is performed on all the target datasets. Binary and multi-classification on all three target datasets are performed in the downstream task. This work also explored label efficiency by performing the experiments on 10, 30, 50, and 100 percent data from the datasets.
The data augmentations are applied to generate two views of a single image which can be further tested for similarity. During the pretext task, flipping, cropping, translation, scaling, grayscale, rotation, blurring, and resizing augmentation techniques are applied to input medical images to obtain better and more generalizable results. Due to the small dataset size, it is required to investigate various finetune scenarios. Only a few primary augmentations, like resizing and cropping, are used during the downstream task to make this approach more compelling. After performing numerous experiments with varied parameter settings, the hyperparameter values that gave promising results during the pretext task are given in Tables \ref{tab2} \& \ref{tab3}. 
Table \ref{tab2} shows the hyperparameters used for binary classification of diabetic retinopathy images for datasets APTOS 2019 and Messidor-I. It shows various probability values used for different augmentation techniques during the pretext task. The batch size used is 128, and the optimizer used for training is LARS (Layer-wise Adaptive Rate Scaling). The initial learning rate used is 0.79, and the weight decay is 10-6. The performance metrics are defined below.
\begin{table}[htbp]
\caption{Hyperparameters for binary classification}
\begin{center}
\begin{tabular}{|c|c|c|}
\hline
\textbf{\textit{Augmentations}}& \textbf{\textit{Parameters}} \\
\hline
\textit{Resize}& \textit{224 X 224}\\
\hline
\textit{Horizontal Flip}& \textit{P=0.5} \\
\hline
\textit{Vertical Flip}& \textit{P=0.5}\\
\hline
\textit{Grayscale}& \textit{P=0.2} \\
\hline
\textit{Gaussian Blur}& \textit{P = 0.5, \newline Kernel size = [21, 21]} \\
\hline
\textit{Batch size}& \textit{128} \\
\hline
\textit{Optimizer}& \textit{LARS} \\
\hline
\textit{Learning Rate}& \textit{0.79} \\
\hline
\textit{Weight-decay} & \textit{$10^{-6}$} \\
\hline
\end{tabular}
\label{tab2}
\end{center}
\end{table}

Table \ref{tab3} shows the hyperparameters used for a multi-classification of diabetic retinopathy images for the dataset APTOS 2019. The augmentations in the multi-classification of DR images are – jitter, affine, and normalization, along with the augmentations used in the binary classification for better performance. The batch size used for multi-classification is 256, and the optimizer used for training is LARS. 

\begin{table}[]
\centering
\caption{Hyperparameters for multi-classification}
\label{tab3}
\resizebox{\columnwidth}{!}{%
\begin{tabular}{|l|c|}
\hline
\textit{\textbf{Augmentations}} & \textit{\textbf{Parameters}}                                                                                                                                                                              \\ \hline
\textit{Resize}                 & \textit{224 X 224}                                                                                                                                                                                        \\ \hline
\textit{Horizontal Flip}        & \textit{P = 0.5}                                                                                                                                                                                          \\ \hline
\textit{Normalization}          & \textit{\begin{tabular}[c]{@{}c@{}}Mean=(0.425, 0.297, 0.212)\\ Standard deviation = (0.276, 0.202, 0.168)\end{tabular}} \\ \hline
\textit{Jitter}                 & \multicolumn{1}{l|}{\textit{Brightness: 0.4; Contrast: 0.4; Saturation: 0.4; Hue: 0.1}}                                                                                                                   \\ \hline
\textit{Affine}                 & \textit{Degrees= (-180, 180), Translate= (0.2, 0.2)}                                                                                                                                                      \\ \hline
\textit{Grayscale}              & \textit{P = 0.2}                                                                                                                                                                                          \\ \hline
\textit{Batch size}             & \textit{256}                                                                                                                                                                                              \\ \hline
\textit{Optimizer}              & \textit{LARS}                                                                                                                                                                                             \\ \hline
\textit{Learning Rate}          & \textit{$10^{-3}$}                                                                                                                                                                                             \\ \hline
\textit{Weight-decay}            & \textit{$5 x 10^{-4}$}                                                                                                                                                                                         \\ \hline
\end{tabular} %
}
\end{table}


The classification task is performed on three datasets – APTOS 2019, Messidor-I, and Fundus Images. Table \ref{tab4} \ shows the results obtained for the binary classification on APTOS 2019 dataset. For APTOS 2019, the proposed method obtains an accuracy of 99.59\%, a precision of 100\%, a recall of 99.54\%, and an F1- score of 99.26\% on only 10\% images. For 100\% images, the accuracy improved by 2.71\%, recall by 5\%, and F1-Score by 3\%.

For another dataset- Messidor-I, the highest accuracy obtained is 98.49\%, precision is 98.65\%, recall is 100\%, and F1 score is 99.99\% as shown in Table \ref{tab5}. 

\begin{table}[]
\caption{Results for binary classification using the proposed approach on APTOS 2019}
\label{tab4}
\resizebox{\columnwidth}{!}{%
\begin{tabular}{|lll|l|l|l|l|}
\hline
\multicolumn{3}{|l|}{\textit{Dataset used}}                                                                                         & {}{}{\textit{Accuracy}} & {}{}{\textit{Precision}} & {}{}{\textit{Recall}} & {}{}{\textit{F1-Score}} \\ \cline{1-3}
\multicolumn{1}{|l|}{\textit{Pretext}}                  & \multicolumn{2}{l|}{\textit{Downstream}}                                  &                                    &                                     &                                  &                                    \\ \hline
\multicolumn{1}{|l|}{{}{}{\textit{EyePACS}}} & \multicolumn{1}{l|}{{}{}{\textit{APTOS 2019}}} & \textit{10\%}  & \textit{96.88}                     & \textit{100}                        & \textit{94.23}                      & \textit{96.44}                        \\ \cline{3-7} 
\multicolumn{1}{|l|}{}                                  & \multicolumn{1}{l|}{}                                     & \textit{20\%}  & \textit{99.48}                     & \textit{100}                        & \textit{98.54}                      & \textit{99.01}                        \\ \cline{3-7} 
\multicolumn{1}{|l|}{}                                  & \multicolumn{1}{l|}{}                                     & \textit{50\%}  & \textit{99.56}                     & \textit{99.01}                         & \textit{99.62}                      & \textit{99.18}                        \\ \cline{3-7} 
\multicolumn{1}{|l|}{}                                  & \multicolumn{1}{l|}{}                                     & \textit{100\%} & \textit{99.59}                     & \textit{100}                        & \textit{99.54}                      & \textit{99.26}                       \\ \hline
\end{tabular}%
}
\end{table}
\begin{table}[]
\centering
\caption{Results for binary classification using the proposed approach on Messidor-I
}
\label{tab5}
\resizebox{\columnwidth}{!}{ %
\begin{tabular}{|llc|c|c|c|c|}
\hline
\multicolumn{3}{|c|}{\textit{\textbf{Dataset used}}}                                                                                 & {\textit{\textbf{Accuracy}}} & {\textit{\textbf{Precision}}} & {\textit{\textbf{Recall}}} & {\textit{\textbf{F1-Score}}} \\ \cline{1-3}
\multicolumn{1}{|l|}{\textit{\textbf{Pretext}}}         & \multicolumn{2}{l|}{\textit{\textbf{Downstream}}}                          &                                             &                                              &                                           &                                             \\ \hline
\multicolumn{1}{|l|}{{\textit{EyePACS}}} & \multicolumn{1}{l|}{{\textit{Messidor-1}}} & \textit{10\%}  & \textit{67.48}                              & \textit{71.43}                                  & \textit{69.89}                               & \textit{72.56}                                 \\ \cline{3-7} 
\multicolumn{1}{|l|}{}                                  & \multicolumn{1}{l|}{}                                     & \textit{20\%}  & \textit{70.31}                              & \textit{75.21}                                  & \textit{74.87}                               & \textit{70.55}                                 \\ \cline{3-7} 
\multicolumn{1}{|l|}{}                                  & \multicolumn{1}{l|}{}                                     & \textit{50\%}  & \textit{74.96}                              & \textit{81.47}                                  & \textit{73.66}                               & \textit{76.75}                                 \\ \cline{3-7} 
\multicolumn{1}{|l|}{}                                  & \multicolumn{1}{l|}{}                                     & \textit{100\%} & \textit{98.49}                              & \textit{98.65}                                  & \textit{100.00}                              & \textit{99.99}                                 \\ \hline
\end{tabular} %
}
\end{table}

For the third dataset- Fundus Images, the accuracy obtained on 100\% images is 98.96\%, precision is 96\%, recall is 99.43\%, and F1 score is 99.67\% as shown in Table \ref{tab6}. 

\begin{table}[htbp]
\centering
\caption{Results for binary classification on Fundus Images
}
\label{tab6}
\resizebox{\columnwidth}{!}{%
\begin{tabular}{|lcc|c|c|c|c|}
\hline
\multicolumn{3}{|c|}{\textit{\textbf{Dataset used}}}                                                                                    & {\textit{\textbf{Accuracy}}} & {\textit{\textbf{Precision}}} & {\textit{\textbf{Recall}}} & {\textit{\textbf{F1-Score}}} \\ \cline{1-3}
\multicolumn{1}{|l|}{\textit{\textbf{Pretext}}}         & \multicolumn{2}{l|}{\textit{\textbf{Downstream}}}                             &                                             &                                              &                                           &                                             \\ \hline
\multicolumn{1}{|l|}{{\textit{EyePACS}}} & \multicolumn{1}{c|}{{\textit{Fundus Images}}} & \textit{10\%}  & \textit{92.15}                              & \textit{93.44}                                  & \textit{91.56}                               & \textit{93.32}                                 \\ \cline{3-7} 
\multicolumn{1}{|l|}{}                                  & \multicolumn{1}{c|}{}                                        & \textit{20\%}  & \textit{93.75}                              & \textit{95.66}                                  & \textit{91.54}                               & \textit{95.98}                                 \\ \cline{3-7} 
\multicolumn{1}{|l|}{}                                  & \multicolumn{1}{c|}{}                                        & \textit{50\%}  & \textit{96.43}                              & \textit{98.32}                                  & \textit{98.22}                               & \textit{96.44}                                 \\ \cline{3-7} 
\multicolumn{1}{|l|}{}                                  & \multicolumn{1}{c|}{}                                        & \textit{100\%} & \textit{98.96}                     & \textit{100.00}                        & \textit{99.43}                      & \textit{99.67}                        \\ \hline
\end{tabular}%
}
\end{table}

Table \ref{ee} shows the results of the multi-classification of DR images by using the same dataset for pretext and downstream tasks, i.e., the Subset of EyePACS.
Table \ref{tab7} shows the outcomes of the three datasets' multi-classification of DR images. The downstream dataset used for the multi-classification of diabetic retinopathy images is APTOS 2019, Messidor-I, and Fundus Images, where the proposed method obtained an accuracy of 83.43\%, 66.39\%, and 91.67\%. The proposed self-supervised learning method outperforms prior state-of-the-art techniques on two datasets - Aptos 2019 and Fundus Images.
\begin{table}[htbp]
\caption{Multi-classification results on eyepacs dataset}
\label{ee}
\resizebox{\columnwidth}{!}{%
\begin{tabular}{|lll|c|c|c|c|}
\hline
\multicolumn{3}{|c|}{\textit{\textbf{Dataset used}}}                                                                             & {}{}{\textit{\textbf{Accuracy}}} & {}{}{\textit{\textbf{Precision}}} & {}{}{\textit{\textbf{Recall}}} & {}{}{\textit{\textbf{F1-Score}}} \\ \cline{1-3}
\multicolumn{1}{|l|}{\textit{\textbf{Pretext}}}         & \multicolumn{2}{l|}{\textit{\textbf{Downstream}}}                      &                                             &                                              &                                           &                                             \\ \hline
\multicolumn{1}{|l|}{{}{}{\textit{Eyepacs}}} & \multicolumn{1}{l|}{{}{}{\textit{Eyepacs}}} & \textit{10\%}  & \textit{74.56}                              & \textit{69.31}                               & \textit{75.32}                            & \textit{70.02}                              \\ \cline{3-7} 
\multicolumn{1}{|l|}{}                                  & \multicolumn{1}{l|}{}                                  & \textit{20\%}  & \textit{77.23}                              & \textit{72.78}                               & \textit{76.99}                            & \textit{73.21}                              \\ \cline{3-7} 
\multicolumn{1}{|l|}{}                                  & \multicolumn{1}{l|}{}                                  & \textit{50\%}  & \textit{77.76}                              & \textit{75.98}                               & \textit{77.06}                            & \textit{74.99}                              \\ \cline{3-7} 
\multicolumn{1}{|l|}{}                                  & \multicolumn{1}{l|}{}                                  & \textit{100\%} & \textit{77.82}                              & \textit{73.71}                               & \textit{77.41}                            & \textit{74.98}                              \\ \hline
\end{tabular}%
}
\end{table}

\begin{figure}[htbp]
\centerline{
\includegraphics[width=8cm, height=7cm]{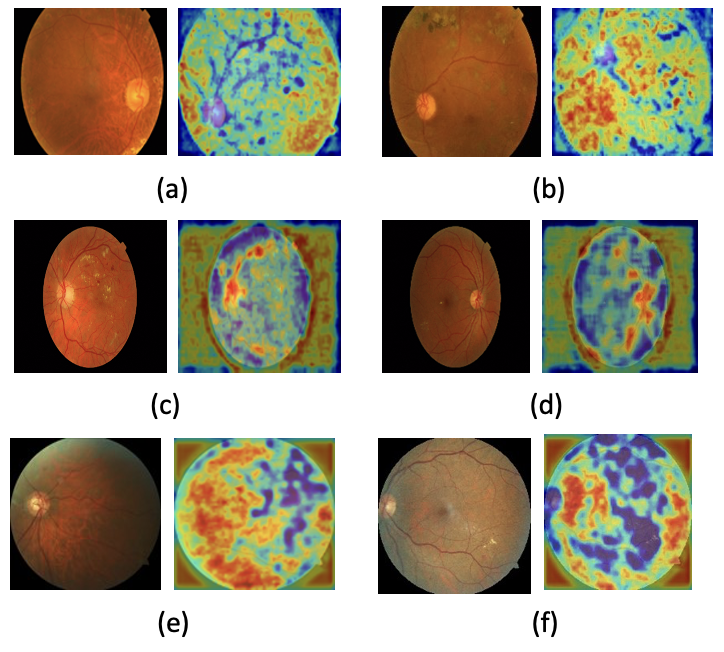} 
}
\caption{ CAMs representations for the datasets used. (a) \& (b) APTOS 2019 (c) \& (d) Messidor I (e) \& (f) Fundus Images
}
\label{fig8}
\end{figure}

Figure \ref{fig8} represents the class activation maps (CAMs) generated for the three downstream datasets.

\begin{table}[]
\caption{Multi-classification results on three downstream datasets
}
\label{tab7}
\resizebox{\columnwidth}{!}{%
\begin{tabular}{|clc|c|c|c|c|}
\hline
\multicolumn{3}{|c|}{\textit{\textbf{Dataset used}}}                                                                                    & {}{}{\textit{\textbf{Accuracy}}} & {}{}{\textit{\textbf{Precision}}} & {}{}{\textit{\textbf{Recall}}} & {}{}{\textit{\textbf{F1-Score}}} \\ \cline{1-3}
\multicolumn{1}{|l|}{\textit{\textbf{Pretext}}}          & \multicolumn{2}{l|}{\textit{\textbf{Downstream}}}                            &                                             &                                              &                                           &                                             \\ \hline
\multicolumn{1}{|c|}{{}{}{\textit{EyePACS}}} & \multicolumn{1}{l|}{{}{}{\textit{APTOS 2019}}}    & \textit{10\%}  & \textit{76.04}                              & \textit{78.45}                               & \textit{89.23}                            & \textit{88.45}                              \\ \cline{3-7} 
\multicolumn{1}{|c|}{}                                   & \multicolumn{1}{l|}{}                                        & \textit{30\%}  & \textit{78.15}                              & \textit{62.32}                               & \textit{83.34}                            & \textit{78.76}                              \\ \cline{3-7} 
\multicolumn{1}{|c|}{}                                   & \multicolumn{1}{l|}{}                                        & \textit{50\%}  & \textit{83.12}                              & \textit{81.12}                               & \textit{82.76}                            & \textit{84.45}                              \\ \cline{3-7} 
\multicolumn{1}{|c|}{}                                   & \multicolumn{1}{l|}{}                                        & \textit{100\%} & \textit{\textbf{83.43}}                     & \textit{81.09}                               & \textit{85.54}                            & \textit{77.86}                              \\ \cline{2-7} 
\multicolumn{1}{|c|}{}                                   & \multicolumn{1}{l|}{{}{}{\textit{Messidor-I}}}    & \textit{10\%}  & \textit{54.16}                              & \textit{67.23}                               & \textit{91.75}                            & \textit{69.34}                              \\ \cline{3-7} 
\multicolumn{1}{|c|}{}                                   & \multicolumn{1}{l|}{}                                        & \textit{30\%}  & \textit{47.22}                              & \textit{54.12}                               & \textit{51.34}                            & \textit{54.97}                              \\ \cline{3-7} 
\multicolumn{1}{|c|}{}                                   & \multicolumn{1}{l|}{}                                        & \textit{50\%}  & \textit{65.83}                              & \textit{64.54}                               & \textit{65.45}                            & \textit{72.21}                              \\ \cline{3-7} 
\multicolumn{1}{|c|}{}                                   & \multicolumn{1}{l|}{}                                        & \textit{100\%} & \textit{\textbf{66.39}}                     & \textit{70.87}                               & \textit{68.79}                            & \textit{71.23}                              \\ \cline{2-7} 
\multicolumn{1}{|c|}{}                                   & \multicolumn{1}{c|}{{}{}{\textit{Fundus Images}}} & \textit{10\%}  & \textit{75.0}                               & \textit{67.98}                               & \textit{67.65}                            & \textit{80.40}                              \\ \cline{3-7} 
\multicolumn{1}{|c|}{}                                   & \multicolumn{1}{c|}{}                                        & \textit{30\%}  & \textit{81.25}                              & \textit{85.90}                               & \textit{91.43}                            & \textit{88.49}                              \\ \cline{3-7} 
\multicolumn{1}{|c|}{}                                   & \multicolumn{1}{c|}{}                                        & \textit{50\%}  & \textit{82.00}                              & \textit{84.12}                               & \textit{80.32}                            & \textit{91.23}                              \\ \cline{3-7} 
\multicolumn{1}{|c|}{}                                   & \multicolumn{1}{c|}{}                                        & \textit{100\%} & \textit{\textbf{91.67}}                     & \textit{86.01}                               & \textit{92.43}                            & \textit{94.54}                              \\ \hline
\end{tabular}%
}
\end{table}

\subsection{Comparison with the existing work}
Domain adaptation-based self-supervised learning on Messidor-I, Fundus Images, and APTOS 2019 datasets is unexplored. This work has considered various methods applied to these datasets for comparison purposes, including supervised learning-based methods. Table \ref{tab9} compares the results obtained from the proposed work with the existing methods for binary classification on Messidor-1, Fundud Images, and APTOS 2019 datasets. 
Chakraborty et al.~\cite{b26} used ANN for binary classification and achieved an accuracy of 97.13\% on the Messidor dataset. A CNN-based model proposed by Islam et al. \cite{b27} for the DR classification APTOS 2019 dataset achieved an accuracy of 98.36\%. However, the proposed method reports improved results (accuracy of 99.59\%, 98.49\%, and 98.96\%) on APTOS 2019, Messidor, and Fundus Images datasets.

\begin{table}[H]
\centering
\caption{Comparative results for binary classification}
\label{tab9}
\resizebox{\columnwidth}{!}{%
\begin{tabular}{|cccc|}
\hline
\multicolumn{1}{|c|}{\textbf{Method}}                    & \multicolumn{1}{c|}{\textbf{Accuracy}} & \multicolumn{1}{c|}{\textbf{Precision}} & \textbf{Recall} \\ \hline
\multicolumn{4}{|c|}{\textbf{Dataset - Messidor}}                                                                                                             \\ \hline
\multicolumn{1}{|c|}{Abramoff et al {[}28{]}}            & \multicolumn{1}{c|}{96.7}              & \multicolumn{1}{c|}{96.80}               & 87.00              \\ \hline
\multicolumn{1}{|c|}{Chakraborty et al {[}26{]}}         & \multicolumn{1}{c|}{97.13}             & \multicolumn{1}{c|}{97.20}               & 97.00              \\ \hline
\multicolumn{1}{|c|}{Dhanasekaran et al. {[}29{]} (SVM)} & \multicolumn{1}{c|}{97.89}             & \multicolumn{1}{c|}{98.68}              & 100.00             \\ \hline
\multicolumn{1}{|c|}{Dhanasekaran et al. {[}29{]} (PNN)} & \multicolumn{1}{c|}{94.76}             & \multicolumn{1}{c|}{96.64}              & 98.46           \\ \hline
\multicolumn{1}{|c|}{Proposed work - SSL Cross domain}   & \multicolumn{1}{c|}{\textbf{98.49}}    & \multicolumn{1}{c|}{98.00}              & \textbf{100.00}    \\ \hline
\multicolumn{4}{|c|}{\textbf{Dataset - APTOS 2019}}                                                                                                           \\ \hline
\multicolumn{1}{|c|}{Islam et al {[}27{]}}               & \multicolumn{1}{c|}{98.36}             & \multicolumn{1}{c|}{98.37}              & 98.36           \\ \hline
\multicolumn{1}{|c|}{Proposed work - SSL Cross domain}   & \multicolumn{1}{c|}{\textbf{99.59}}    & \multicolumn{1}{c|}{\textbf{100.00}}       & \textbf{99.00}  \\ \hline
\end{tabular} %
}

\end{table}

Table \ref{tab10} displays the results of a comparison between the proposed study and previous work in multi-class classification of DR images. Kassani et al. [32] reported the highest accuracy for multi-class classification was 83.09\%. The remaining works reported an accuracy below 75\% for classifying diabetic retinopathy images. 

\begin{table}[H]
\centering
\caption{Comparative results for multi-classification on APTOS 2019 dataset. \textit{No suitable previous work found for other datasets for multi-class classification}}
\label{tab10}
\resizebox{\columnwidth}{!}{%
\begin{tabular}{|c|c|c|c|}
\hline
Authors                          & Accuracy       & Precision      & Recall      \\ \hline
Kassani et al {[}32{]}           & 83.09          & \textbf{88.24} & 82.35       \\ \hline
Gangwar \& Ravi {[}33{]}         & 72.33          & -              & -           \\ \hline
Proposed Work - SSL Cross domain & \textbf{83.43} & 81.00             & \textbf{85.00} \\ \hline
\end{tabular} %
}
\end{table}

The comparisons in Tables \ref{tab9} and \ref{tab10} suggest that the performance achieved with the proposed work is improved over previous works for binary and multi-classification of diabetic retinopathy images. 

\subsection{Label efficiency in cross-domain knowledge transfer}
\begin{figure}[htbp]
\centerline{
\includegraphics[width=8cm, height=4cm]{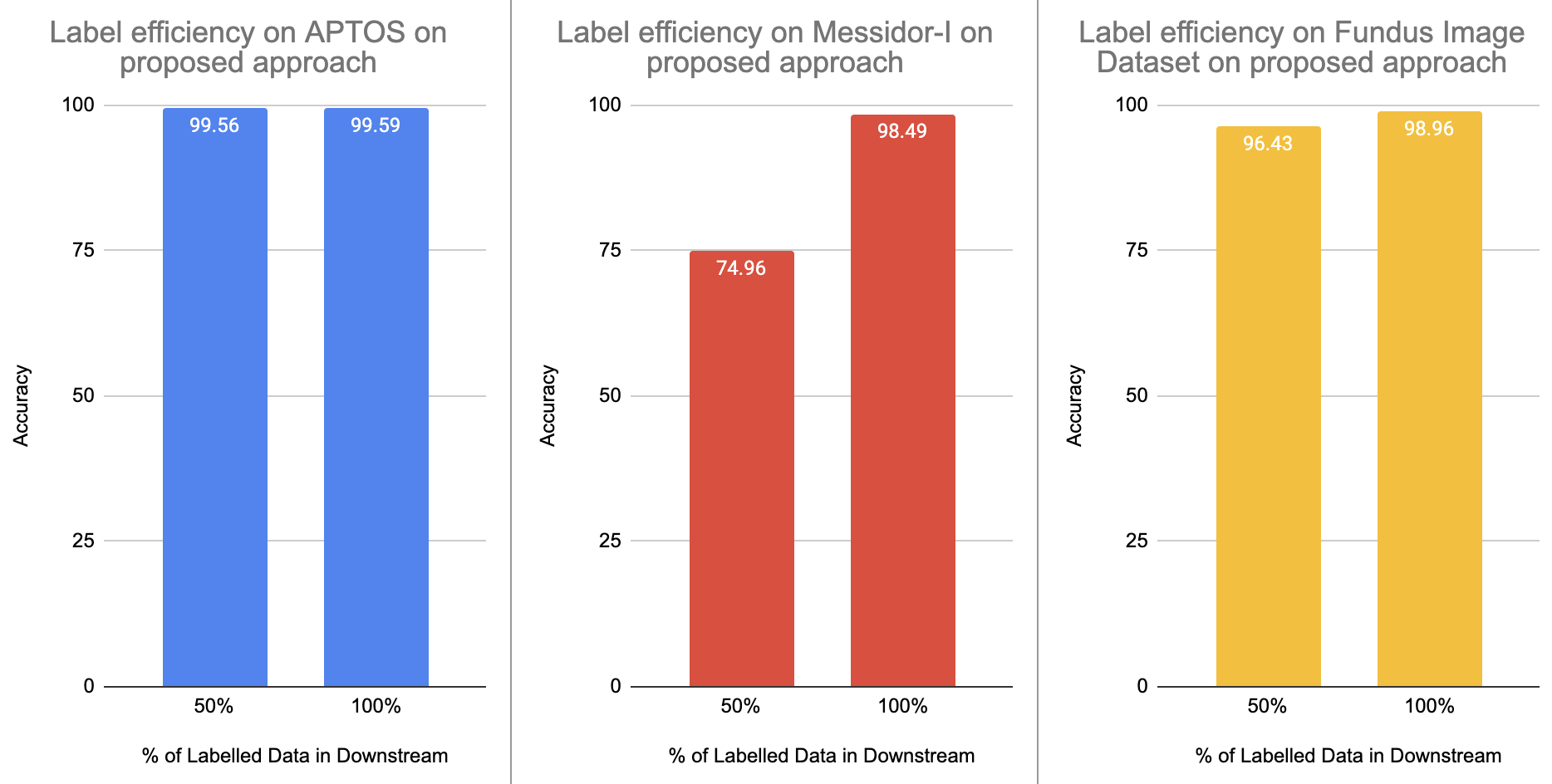} 
}
\caption{Label efficiency on binary classification tasks for all three target datasets, which shows cross-domain knowledge transfer achieves comparable performance with only $50\%$ label being used against the fully supervised model.
}
\label{figbg}
\end{figure}

\begin{figure}[htbp]
\centerline{
\includegraphics[width=8cm, height=4cm]{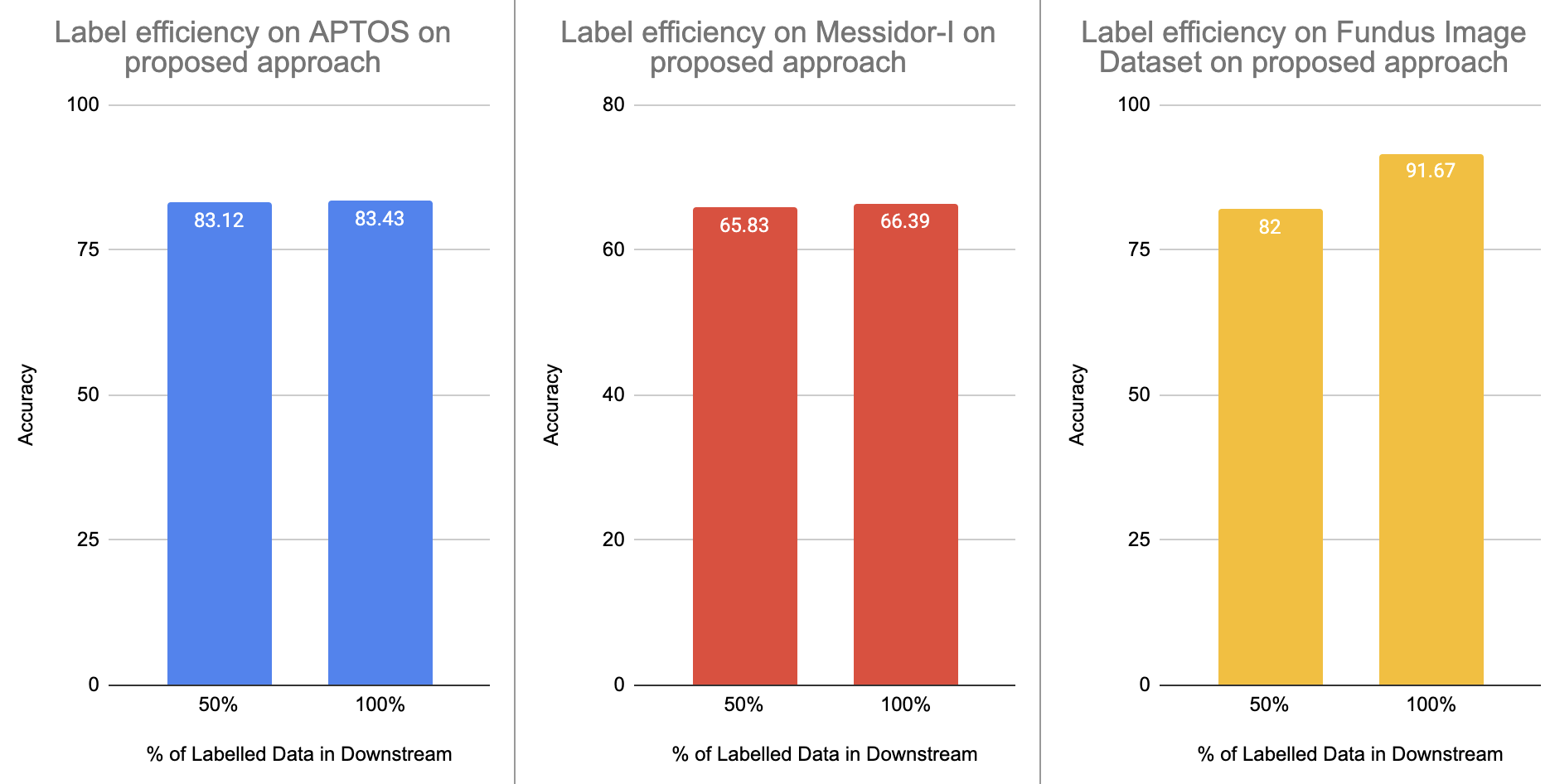} 
}
\caption{ Label efficiency on multi-class classification tasks for all three target datasets, which shows cross-domain knowledge transfer achieves comparable performance with only $50\%$ label being used against the fully supervised model.
}
\label{figmg}
\end{figure}

The proposed self-supervised cross-domain knowledge method obtains concrete evidence for label efficiency. Result comparisons on both downstream tasks show that model achieves comparable performance when only $50\%$ labels (half supervised) are used against fully supervised models with $100\%$ labels. It is observed that downstream task performance differences between partially supervised and fully supervised are in close on at-least two datasets out of three target datasets for both classification tasks. Further, it is noticeable that the training portion of all the target datasets consists of only a few labeled examples in the range of $500$ to $2500$. Label efficiency is illustrated in the figures \ref{figbg} \& \ref{figmg}.

\section{Conclusion}
This work proposes a label-efficient self-supervised representation learning-based method for diabetic retinopathy image classification in cross-domain settings. The proposed work has been evaluated qualitatively and quantitatively on the publicly available EyePACS, APTOS 2019, MESSIDOR-I, and Fundus Images datasets for binary and multi-classification of DR images. The qualitative evaluation shows that the proposed approach learns explainable image representations.  Moreover, the proposed approach uses only a few training samples for training and outperforms the existing DR image classification methods, even in cross-domain settings. In future work, the proposed approach can be used to investigate other downstream tasks, such as segmentation and localization. Further, non-contrastive methods for representation learning can be examined to perform downstream tasks on DR images in cross-domain settings.

\vspace{12pt}

\end{document}